\documentclass[twocolumn,showpacs,preprintnumbers,amsmath,amssymb]{revtex4-1}
\usepackage{mathrsfs}
\usepackage{txfonts}
\usepackage{amsmath}
\usepackage{stmaryrd}
\usepackage{amssymb}
\usepackage{setspace}  
\usepackage{array}
\usepackage{multirow}

\usepackage{graphicx}
\usepackage{dcolumn}
\usepackage{bm}

\begin{document}

\title{Long-range and frustrated spin-spin interactions in crystals of cold polar molecules}

\author{Y. L. Zhou$^1$, M. Ortner$^{2,3}$, P. Rabl$^{3}$}

\affiliation{$^1$College of Science, National University of Defense Technology, Changsha, 410073, China}
\affiliation{$^2$Institute for Theoretical Physics, University of Innsbruck, 6020 Innsbruck, Austria}
\affiliation{$^3$Institute for Quantum Optics and Quantum Information of the Austrian Academy of Sciences, 6020 Innsbruck, Austria}


\date{\today}

\begin{abstract}
We describe a simple scheme for the implementation and control of effective spin-spin interactions in self-assembled crystals of cold polar molecules. In our scheme spin states are encoded in two long-lived rotational states of the molecules and coupled via state dependent dipole-dipole forces to the lattice vibrations. We show that by choosing an appropriate time dependent modulation of the induced dipole moments the resulting phonon-mediated interactions compete with the direct dipole-dipole coupling and lead to long-range and tunable spin-spin interaction patterns. We illustrate how this technique can be used for the generation of multi-particle entangled spin states and the implementation of spin models with longe-range and frustrated interactions which exhibit non-trivial  phases of magnetic ordering.
\end{abstract}



\maketitle

\section{Introduction}


The quantum Ising model describes a system of interacting spins where the coupling among the spins competes with a  transverse magnetic field. It is one of the simplest models which captures many essential aspects of quantum magnetism and has been successfully applied to study the transition from a paramagnetic phase to ferro- or anti-ferromagnetic ordering \cite{quantum-phase-book}. Due to its practical relevance this model has been widely studied and for many realizations with nearest neighbor interactions the ground and thermal states predicted by this model are well understood.   However, less is known about more general Ising models with long-range or frustrated spin-spin interactions~\cite{frustrated-book}, where already identifying the ground state properties can be a complicated and numerically demanding task. Apart from its important role in condensed matter physics the Ising and related spin models have recently also attracted a lot of attention in the field of quantum information processing. Here the dynamics generated by a controlled Ising interactions between qubits can be used to generate various two and multi-partite entangled states as a fundamental resource for quantum computation \cite{cluster1,cluster2,graph1}. These potential applications have over the past years stimualted a lot of additional work in the field of spin models focused in particular on the out of equilibrium dynamics and entanglement properties of this system~\cite{ent-frus}.

The broad interest in a more detailed understanding and control of interacting spin systems on a small and larger scale has stimulated various proposals for the implementation and simulation of quantum Ising models using isolated atomic or coherent solid state systems~\cite{qsor}. A pioneering role in this context is played by systems of trapped ions~\cite{HaeffnerPhysRep2008,WinelandLaserPhysics2011}, where effective, phonon-mediated spin-spin interactions can be implemented and controlled by applying state-dependent light forces~\cite{MM gate,ions-spin, XXZmodel, IsingModels,deng,Zhu,Lin}. Indeed, several proof-of-principle experiments~\cite{Friedenauer,Kim-prl,exp,qs frustrated ising} have already demonstrated the possibility to simulate Ising interactions with up to nine spins and with such ion trap quantum simulators it might soon be possible to outperform the best numerical simulations which are achievable on classical computers today. Based on recent advances in cooling and trapping techniques for diatomic molecules~\cite{coolingEXP1,coolingEXP2,coolingEXP3,coolingEXP4,coolingEXP5,coolingEXP6,coolingEXP7,coolingEXP8}, it is expected that in the near future a similar level of control can be achieved with ensembles of ultra-cold polar molecules.  In this system spin states can be encoded in long-lived rotational or hyperfine states of the molecules, which can then be manipulated with microwave fields and coupled via strong  electric dipole-dipole interactions~\cite{revPM1,revPM2}.  Compared to ions, polar molecules can easily be trapped in standard optical lattice potentials of different geometries or -- by aligning their dipole moments -- be stabilized in a high density crystalline phase~\cite{MDC1,MDC2,MDC3,rabl-pra}.  The combination of these exceptional properties make polar molecules one of the most promising systems for large scale simulation of non-trivial quantum spin models~\cite{Micheli-Nat,Micheli2-Nat,schachenmayer,HerreraPRA,TrefzgerNJP2010,WallPRA2010,KestnerPRB2010,Gorshkov}.

In a recent work \cite{michael} we have analyzed the implementation of spin-spin interactions in a system of polar molecules prepared in a self-assembled dipolar crystal under 1D or 2D trapping conditions. We have shown that in this setting resonantly enhanced phonon-meditated spin-spin interactions can dominate over the direct couplings and we proposed to use this feature for the implementation of local qubit operations in  molecular quantum computing schemes~\cite{DeMille,QIPswitch}.  In this paper we extend this analysis and investigate potential applications of this technique for the design of more general Ising models with non-trivial spin-spin interaction patterns.  In particular, we show that in close analogy to trapped ion system we can use a time-dependent modulation of the induced dipole moments to address collective phonon modes in the crystal. The  competition between short-range direct and long-range phonon-mediated spin-spin interactions then allows us to tailor the resulting effective spin-spin couplings and tune the interaction strength and signs freely via changing the detuning between the dipole moment and phonon frequencies.  We illustrate how this technique can be used in the case of a small 1D crystal of polar molecules to implement frustrated spin-spin couplings and discuss applications for the generation of multi-particle entangled states.  Further, we provide an outlook, how this technique could be extended for simulation of larger spin systems.

The remainder of the paper is organized as follows. In Sec. II we introduce our model and present a general derivation of the resulting Ising Hamiltonian. In Sec. III we discuss  the validity of our spin model and how it can be used to generate a graph state using a simple toy system of three molecules in a harmonic trap. In Sec. IV we use the same setting to illustrate the frustrated character of this Ising Hamiltonian and describe the accessible `phases' and the entanglement properties of the ground states. In Sec. V we discuss potential generalization of this techniques to larger spin systems and finally in Sec. VI we present more details on the encoding of qubits states in rotational states of the molecules. The main results and conclusions of this work are summarized in Sec. VII.

\begin{figure}
\includegraphics[scale=0.5]{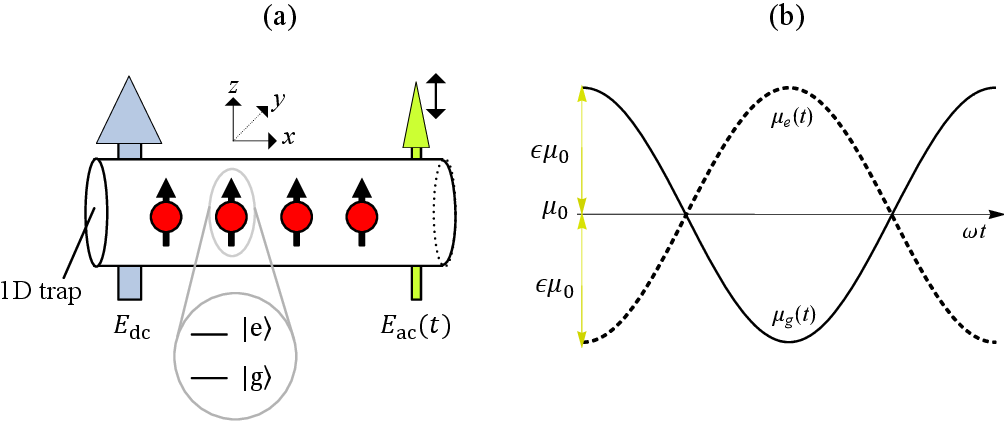}
\caption{(Color online) Setup. (a) Polar molecules with two internal states $|g\rangle$ and $|e\rangle$ are confined in a 1D tube by a strong transverse trapping potential. An electric dc field $E_{\mathrm{dc}}$ is used
 to align the dipole moment of the polar molecules while an additional weaker ac field
 $E_{\mathrm{ac}}(t)$ is used to induce a time dependent modulation of dipole moment,  which is of opposite sign for the two states  as shown in (b). See text for more details.}
 \label{fig:setup}
\end{figure}

\section{Spin-spin interactions in molecular dipolar crystals}
We consider a set of $N$ polar molecules with dipole moments aligned along the $z$-axis by an external bias field $\vec E_{b}=E_b \vec e_z$ and their motion confined to the $x,y$ plane by a strong optical or electric trapping potential.  For simplicity we will focus in the following mainly on a one-dimensional (1D) trapping configuration as shown in Fig.~\ref{fig:setup}, but our results can be generalized to 2D or intermediate trapping conditions. For molecules of mass $m$ the system is then described by the Hamiltonian
\begin{eqnarray}\label{Htot}
H=\sum_{i}\left(\frac{p_i^2}{2m} +\frac{1}{2}m\nu^2x_i^2\right)+V_{dd}(\{x_i\}),
\end{eqnarray}
where $x_i$ and $p_i$ are the position and momentum operators of the molecules and $\nu$ denotes the trapping frequency of an additional weak confinement potential along the tube.
The dipole-dipole interaction is
\begin{equation}
V_{dd}(\{x_i\})= \frac{1}{8\pi\epsilon_0}Ê\sum_{i\neq j}   \frac{\vec \mu_i \vec  \mu_j- 3 (\vec \mu_i \vec e_x)(\vec \mu_j \vec e_x)}{|x_i-x_j|^3},
\end{equation}
where the $\vec \mu_i$ is the dipole operator of the $i$-th molecule.

In the following we chose two long-lived internal states of the molecules to encode our effective spin states and denote them  by $|g\rangle$ and $|e\rangle$. The corresponding induced dipole moments $\langle g|\vec \mu|g\rangle=  \mu_g \vec e_z$ and $\langle  e|\vec \mu|e\rangle=  \mu_e \vec e_z$ are a function of the applied bias field $E_b$ and we assume that the dipole moments satisfy the following conditions: i) $\mu_e(E_0)=\mu_g(E_0)=\mu_0$ for a specific value of the bias field $E_0$, ii)  $\partial_E \mu_e|_{E_0}=-\partial_E \mu_g|_{E_0}$ and iii) $|\langle e |\vec \mu | g\rangle |\ll \mu_0$.  In Sec. VI below we will describe in more detail how these conditions can be achieved for a specific set of rotational states and related ideas have been discussed in Ref.~\cite{michael,schachenmayer,Gorshkov}.  Under the validity of assumptions i)-iii) the combination $E_b(t)=E_{0}+E_{\mathrm{ac}}\cos(\omega t)$ of a static and a weak oscillating bias field then allows us to implement a dipole operator of the form $\vec \mu=\mu_0 \vec e_z (1+\epsilon\cos(\omega t)\sigma^z)$, where $\epsilon\ll1$ and $\sigma^z=|g\rangle\langle g|-|e\rangle\langle e|$ is the Pauli operator.
%
%
We write the dipole-dipole interaction as a sum of two parts, $V_{dd}(x_i-x_j)=V_{dd}^0+V_{dd}^1(t)$ where  for $D=\mu_0^2/(4\pi\epsilon_0)$ the static and state independent part is given by
\begin{equation}\label{sub1}
V_{dd}^0=\frac{D}{2} \sum_{i\neq j}\frac{1}{|x_i-x_j|^3},
\end{equation}
while the remaining part
\begin{equation}
V_{dd}^1(t)=\frac{D}{2} \sum_{i\neq j}\frac{2\epsilon\sigma_i^z\cos(\omega t)+\epsilon^2\sigma_i^z\sigma_j^z\cos^2(\omega t) }{|x_i- x_j|^3},
\end{equation}
depends on the oscillating dipole moments and couples internal and external degrees of freedom.

\subsection{Spin-phonon interactions in a dipolar crystal}
In the limit $\epsilon\ll1$ the dipole-dipole coupling is dominated by the state independent part
$V_{dd}^0$ given in Eq.~\eqref{sub1} which in analogy to the Coulomb interaction in  trapped ion systems \cite{HaeffnerPhysRep2008,WinelandLaserPhysics2011} stabilizes the molecules against close encounter collisions and leads to the formation of a dipolar (quasi) crystal at low temperatures~\cite{MDC1,MDC2,MDC3,rabl-pra}. In this limit molecules become localized at equilibrium positions $x^0_i$ and up to second order in the remaining small fluctuations $\delta x_i$ the crystal dynamics can be described by the phonon Hamiltonian  $H_p\equiv H(\epsilon\rightarrow0)=  \sum_n\hbar \omega_n a_n^{\dag}a_n$ \cite{rabl-pra,michael-njp}. Here   $a_n^{\dag}(a_n)$ and $\omega_n$ are the creation (annihilation) operators and phonon frequency of the $n$th mode respectively, and the total Hamiltonian~\eqref{Htot} can now be written as
\begin{equation}
H= H_p + V_{dd}^1(t).
\end{equation}
In contrast to $V_{dd}^0\propto\mu_0^2$, the term $V_{dd}^1(t)\propto(\epsilon\mu_0^2,\epsilon^2\mu_0^2)$ contains spin-dependent dipole-dipole interaction, but is reduced by  $\epsilon\ll 1$. Therefore, we can expand $V_{dd}^1(t)$ in terms of the small parameters $\epsilon$ and $|\delta x_i|/a$, where $a$ is the typical lattice spacing. To the lowest relevant order in these two parameters we obtain~\cite{rabl-pra,michael,herrera-arXiv}
\begin{equation}\label{V1app}
V_{dd}^1(t)\approx\sum_{i\neq j}\frac{\hbar G_{ij}^0}{2}\sigma_i^z\sigma_j^z\cos^2(\omega t)+\sum_{n,i}\hbar g_{n,i}\sigma_i^z(a_n+a_n^{\dag})\cos(\omega t),
\end{equation}
where 
\begin{eqnarray}
\hbar G_{ij}^0&=&\frac{D\epsilon^2}{ |x_i^0-x_j^0|^3},\\
\hbar g_{n,i}&=&-\sqrt{\frac{\hbar}{2m\omega_n}}\sum_{j\neq i}\frac{3D\epsilon(x_i^0-x_j^0)}{|x_i^0-x_j^0|^5}(c_{n,i}-c_{n,j}).
\end{eqnarray}
Here the $c_{n,i}$ are the normalized mode function amplitudes defined by $\delta x_i=\sum_n\sqrt{\hbar/(2m\omega_n)}c_{n,i}(a_n+a_n^{\dagger})$ and in Eq.~\eqref{V1app} we have already neglected an oscillating single spin term, which does not give a relevant contribution in the dynamics discussed below. The first term in Eq.~\eqref{V1app} is the direct spin-spin interaction and the last term couples the internal states of the polar molecules with their external motion. In the interaction picture with respect to $H_{p}$, the total Hamiltonian of the system  simplifies to
\begin{eqnarray}\label{eq:Hrotframe}
H(t)&=&\sum_{i\neq j}\frac{\hbar G_{ij}^0}{2}\sigma_i^z\sigma_j^z\cos(\omega t)^2\nonumber\\
&&+\sum_{n,i}\hbar g_{n,i}\cos(\omega t)\sigma_i^z(a_n e^{-i\omega_nt}+a_n^\dag e^{i\omega_nt}).
\end{eqnarray}
If we assume that the frequency of the dipole moment is not resonant with any phonon mode i.e., the condition $|\Delta_n|=|\omega-\omega_n|\gg g_{n,i},\ \forall i,n$ is satisfied, the excitations of real phonons can be avoided~\cite{MM gate,ions-spin,IsingModels,deng,Zhu,Lin,atom-spin, James,michael}. On a timescale which is long compared to $|\Delta_n|^{-1}$ the phonon degrees of freedom can be eliminated and the remaining slowly spin dynamics can be described by an effective Hamiltonian (see App.~\ref{app})
\begin{equation}\label{eq:Heff}
H_{\rm eff}=\sum_{i<j}\hbar G_{ij}\sigma_i^z\sigma_j^z,
\end{equation}
where $G_{ij}=(G_{ij}^0+G_{ij}^1)/2$ and
\begin{eqnarray}
G_{ij}^1=\sum_n\frac{2g_{n,i}g_{n,j}\omega_n}{(\omega^2-\omega_n^2)}.
\end{eqnarray}
Note that in Eq.~\eqref{eq:Heff} the bare spin-spin coupling $G_{ij}^0\cos^2(\omega t)$ has been averaged to $G_{ij}^0/2$.

We see that in the effective Ising model in Eq.~\eqref{eq:Heff}  the phonon-mediated coupling $G^1_{ij}$ is added to the bare dipole-dipole interaction $G^0_{ij}$ which depending on the phonon structure and the modulation frequency $\omega$ can in general lead to long-range and more complicated spin-spin interaction patterns.

\section{Toy system: Three molecules in a harmonic trap}
As a first example we will show in this section, how the general approach outlined above can be used to create a specific spin entangled state, namely a so-called graph state, for the simplest case of three molecules in a harmonic trap.  We use this example in particular also to discuss the achievable fidelities of such a state and the validity of our effective spin model.

\subsection{Harmonic trap}

We consider a set of $N$ polar molecules which are trapped in a harmonic potential with trapping frequency $\nu$, such that the classical equilibrium positions $x_i^0$ are determined by the condition
\begin{equation}
m\nu^2 x_i^0 - 3D\sum_{j\neq i}  \frac{(x_i^0-x_j^0)}{|x_i^0-x_j^0|^5} =0,
\end{equation}
and the distance between particles is in general not constant.  We denote by $a={\rm min} \{ |x_i^0-x_j^0|\}$ the typical lattice spacing at the center of the trap  which we write as $a=\xi_N\sqrt[5]{D/(m\nu^2)}$, where $\xi_N$ is a numerical constant which is $\xi_2=\sqrt[5]{6}$ for two molecules, $\xi_3\simeq 1.26$ for three molecules and scales as $\xi_N\sim N^{-2/5}$ for larger $N$~\cite{rabl-pra}.  By introducing dimensionless positions $\bar{x_i}=x_i/a$, the
 couplings normalized to the relevant dipole-dipole interaction energy can then be written as
%
$\bar G_{ij}^0=\hbar G_{ij}^0/(D\epsilon^2/a^3)=1/|\bar x_i^0-\bar x_j^0|^3$ and $\bar G_{ij}^1=\hbar G_{ij}^1/(D\epsilon^2/a^3)=\sum_n   \bar g_{n,i}\bar g_{n,j}/(\bar\omega^2-\bar\omega_n^2)$ where $\bar\omega_{n}=\omega_{n}/\nu$  and
\begin{equation}
\bar g_{n,i}=-\frac{3}{\xi_N^{5/2}}\sum_{j\neq i}\frac{(\bar x_i^0-\bar x_j^0)}{|\bar x_i^0-\bar x_j^0|^5}(c_{n,i}-c_{n,j}).
\end{equation}
\begin{figure}
\includegraphics[width=0.4\textwidth]{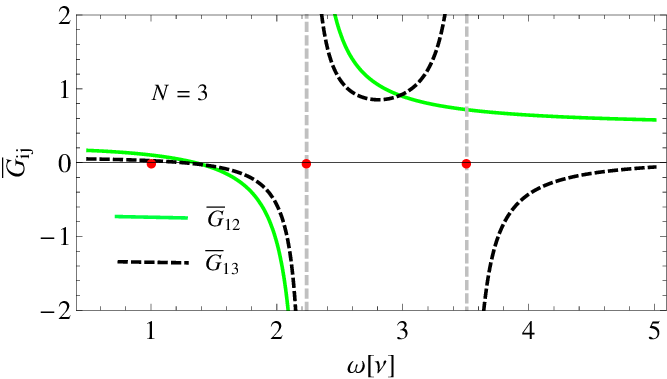}
\caption{(Color online) The normalized spin-spin coupling strength $\bar G_{ij}$ for three polar molecules in a 1D harmonic trap as a function of the dipole moment oscillation frequency $\omega$. The normal mode frequencies are indicated by the solid dots in the \emph{x}-axis. }
\label{fig:Couplings}
\end{figure}
In Fig.~\ref{fig:Couplings} we  plot the total spin-spin interaction strength $\bar G_{ij}$ for $N=3$ as a function of the dipole modulation frequency $\omega$. When we tune $\omega$ away from a specific phonon frequency the phonon mediated spin-spin coupling contains contributions from all modes except the center-of-mass (COM) mode where all $c_{n,i}=1/\sqrt N$ are the same.   It shows that we can control the coupling strength and sign completely by varying the detuning. One specific mode will dominate $\bar G_{ij}^1$ when we tune $\omega$ close to resonance. Since such an vibrational eigenmode is shared by all the particles this situation leads to long-range interactions and provides us with the opportunities to generate highly entangled spin states \cite{ent-frus}. In the opposite case of far detuning from all the normal modes, phonon-mediated spin-spin couplings will be strongly suppressed, and thereby the bare, almost nearest-neighbor interactions dominate.

\subsection{Generation of graph states}

In the following we illustrate how the effective spin-spin interaction entangles particles by presenting a numerical simulation for the case of three trapped polar molecules. As shown in Fig.~\ref{fig:Couplings}, for  $\omega=2.977\nu$ all the couplings between different particles are the same, i.e., $G_{12}=G_{23}=G_{13}=0.911D\epsilon^2/(\hbar a^3)$. This corresponds to the Ising Hamiltonian $H=\sum_{i,j}\hbar G\sigma_i^z\sigma_j^z$, which can directly be used for generation of a three-qubit graph state \cite{cluster1, cluster2, graph1}
\begin{equation}\label{eq:graph}
|\Phi_g\rangle=\frac{1}{2^{3/2}}(|g\rangle_1+|e\rangle_1\sigma_2^z)
\otimes(|g\rangle_2+|e\rangle_2\sigma_3^z)\otimes(|g\rangle_3+|e\rangle_3\sigma_1^z),
\end{equation}
by evolving an initial state $(|g\rangle+|e\rangle)^{\otimes 3}/2^{3/2}$ for a time $t=\pi/(4 G_{12})$.


To see how the graph state can be generated and also to confirm the validity of our model, we compare the effective spin evolution of Eq.~\eqref{eq:Heff} with the dynamics of the full time dependent Hamiltonian~\eqref{eq:Hrotframe},  which is evaluated in App.~\ref{app}.  Fig. 3(a) shows the evolution of the graph state fidelity $P_{\rm graph}(t)=\mathrm{Tr}_p\{\rho(t) |\Phi_g\rangle\langle \Phi_g|\}$ for $\epsilon=0.05$ and $\epsilon=0.1$ (solid lines), where $\rho(t)$ is the density matrix evolved under the full Hamiltonian~\eqref{eq:Hrotframe} and $\mathrm{Tr}_p$ denotes the partial trace over the phonon degrees of freedom. The dashed line shows the evolution calculated from the effective spin Hamiltonian ~\eqref{eq:Heff}. The plot clearly shows that our effective spin Hamiltonian agrees well with the full time dependent model and when $G_{12}t =\pi/4$ three molecules are in the graph state with a fidelity of $96.5\%$ in the case $\epsilon=0.1$. For molecular dipole moments of a few  Debye and lattice spacing of a few hundred nanometers the typical dipole-dipole interaction strength can be tens of kHz, in principle limited by the transverse optical trapping frequency only.  If we set $\epsilon=0.1$ we obtain $G_{ij}/(2\pi)\sim 1$ kHz and the typical time scale for the generation of the graph state is about 150 $\mu$s. Larger interactions could be achievable in electrostatic traps~\cite{AndreNatPhys2006}, where a stronger transverse confinement would allow dipole-dipole couplings $D/a^3$ in the MHz regime~\cite{rabl-pra}.

\begin{figure}
\begin{center}
\includegraphics[width=0.40\textwidth]{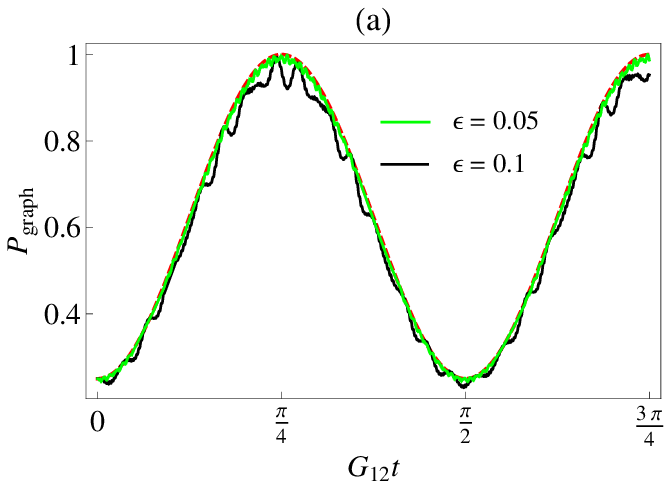}
\includegraphics[width=0.40\textwidth]{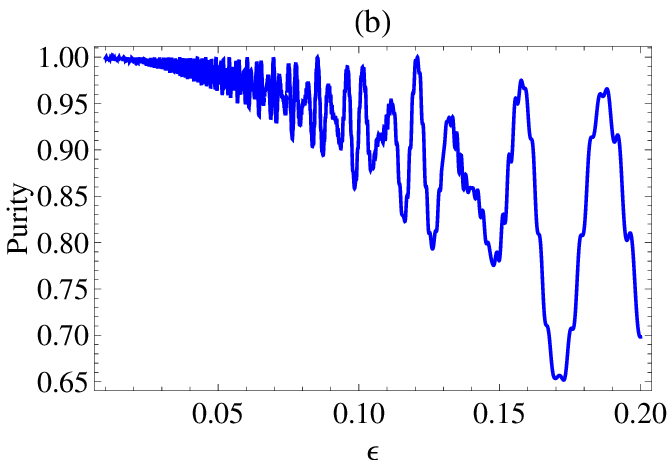}
\caption{(Color online) (a) The occupation probability $P_{\rm graph}$ of graph state~\eqref{eq:graph} calculated with the full Hamiltonian (dashed red line) and the effective spin Hamiltonian with $\epsilon=0.05$ (solid green line) and 0.1 (solid blue line), respectively.  (b) The purity of the reduced state of the effective spin model as a function of $\epsilon$ at time $t=\pi/(4G_{12})$. In both plots we have assumed $\omega=2.977\nu$ such that the nearest-neighbor interaction equal to the next-nearest-neighbor interaction, $G_{12}=G_{13}$.}
\label{fig:fideltiy}
\end{center}
\end{figure}

\begin{table*}
\renewcommand{\arraystretch}{1.5}
\begin{centering}
\begin{tabular}{|c|c|c|c|c|}
\hline
\,\,\, region \,\,\, & \,\,\, coupling strength \,\,\,  & \,\,\, ground state \,\,\,  & \,\,\, spin order \,\,\, & \,\,\, transition \,\,\, \\
 \hline
Ia & \,\,\,$G_{12}>0,\, G_{12}>G_{13}$\,\,\, & $|ege\rangle,\, |geg\rangle$  & AFMS  & \multirow{2}{*}{AFMS $\leftrightarrow$ FM} \\
\cline{2-4}
Ib & $G_{12}, G_{13}<0 $ & $|eee\rangle,\, |ggg\rangle$  & FM  & \\
\hline
\multirow{2}{*}{II} & $G_{12}>G_{13}>0$ & $|ege\rangle,\, |geg\rangle$  & AFMS  & \multirow{2}{*}{\,\,\, AFMS $\leftrightarrow$ AFMA \,\,\,} \\
\cline{2-4}
& $G_{13}>G_{12}>0$ & \,\,\, $|eeg\rangle,\, |gee\rangle,\, |egg\rangle,\, |gge\rangle$  \,\,\, & AFMA  &  \\
\hline
III & $G_{12}>0,\, G_{13}<0$ & $|ege\rangle,\, |geg\rangle$  & AFMS  & \\
\hline
\end{tabular}
\caption{Summary of the spin order of the ground state of the three-spin system in the limit $B^x\rightarrow 0$ and for the regions I, II and III of the modulation frequency $\omega$ as described in the text.}
\label{tab:spin-order}
\end{centering}
\end{table*}

Discrepancies between the exact results and the effective Hamiltonian arise from higher order terms in the expansion parameters, fast oscillating terms which we have omitted in the effective spin dynamics and the possibility of phonons being excited. The small high frequency oscillations superimposed on the solid curve come from off-resonant couplings between the spin and phonons and the fast oscillating bare coupling term $\cos(2\omega t)\sigma^z\sigma^z$. To study the parasitic entanglement between spins and phonons we plot in
Fig.~\ref{fig:fideltiy}(b) the purity of the reduced density matrix of the spin chain $P_s=\mathrm{Tr}[\rho_s^2]$ for the final time $t=\pi/(4G_{12})$ where $\rho_s=\mathrm{Tr}_p[\rho]$ is the reduced spin density operator. The purity decays with $\epsilon^2$, but specific values where all phonon modes rephase (maxima of $P_s$) can be exploited to achieve high fidelities for not too small coupling. In this plot we have assumed the the phonon modes are initially in the ground state. Actually, because the coupling between spins and phonons is off-resonant, the phonon-mediated spin-spin coupling is robust to thermal excitation of those modes \cite{MM gate}, but for larger temperatures the fidelity will decay as $\sim \epsilon^2 n_{th}$ where $n_{th}\approx k_BT/\hbar \nu$ is the characteristic phonon occupation number.

From our analysis we conclude that our effective Ising model given in Eq.~\eqref{eq:Heff} is a valid approximation to the full system dynamics under the relevant conditions. This Ising model with long-range coupling can be used to generate highly entangled states, which are almost decoupled from the external motion of the molecules. Although cooling polar molecules is more difficult than ions, temperatures of $T\sim$ mK and below have been reached in several recent experiments~\cite{coolingEXP3,coolingEXP4,coolingEXP5,coolingEXP6,coolingEXP7,coolingEXP8}. This temperature is sufficient to avoid unwanted thermal effects on the effective spin-spin couplings \cite{loewen}.

\section{Frustrated spin-spin interactions }
In the last section we have discussed how to realize $\sigma^z\sigma^z$ interactions with different spatial profiles through exchanging virtual phonons. We now consider a simple extension of this model by adding a transverse field $B^x$,
\begin{equation}\label{eq:HIsing}
H=\hbar B^x\sum_i\sigma_i^x+\sum_{i<j}\hbar G_{ij}\sigma_i^z\sigma_j^z.
\end{equation}
This effective transverse field can be implemented  by coupling the states $|g\rangle$ and $|e\rangle$ either directly via a resonant microwave field or via optical Raman transitions.
For a homogenous system with nearest neighbor couplings the Ising model in Eq.~\eqref{eq:HIsing} exhibits a well known transition from a paramagnetic to a ferromagnetic (FM) or antiferromagnetic (AFM) phase, depending on the sign of the $G_{i,i+1}$~\cite{quantum-phase-book}. For long-range spin-spin interactions the ground state of the Ising model is in general more involved as, for example, the competition between nearest and next-nearest neighbor coupling can lead to frustration effects where for $B_x\rightarrow 0$ the energy is minimized by a superpositions of multiple degenerate spin configurations~\cite{frustrated-book}.


\subsection{Frustrated three-spin models with polar molecules}

Let us continue with the three spin system described above, which is the minimal setting where frustration effects can occur~\cite{exp}.  As mentioned above for $B^x\rightarrow 0$ and $G_{ij} <0$ the energy is always minimized when all spins are aligned, i.e., $|ggg\rangle$ and $|eee\rangle$ and the ground state is ferromagnetic.  For positive couplings the spins have to be anti-aligned to lower the energy and the ground state is AFM. It is possible to get the optimal configuration for a 1D spin array which just has nearest neighboring interactions. But for a three-spin-array with long-range interactions, which can be seen as a 2D triangular lattice, it is impossible to arrange the spins in a way such that each spin is anti-parallel to all its interacting partners. The system then does not have a simple periodic ground state  and is said to be frustrated \cite{frustrated-book}. For example, in our system when $G_{12}$ or $G_{13}>0$, the system is frustrated and the Ising ground state is an entangled superposition of six AFM states~\cite{exp}: four antiferromagnetic asymmetric (AFMA) states $|gge\rangle, |eeg\rangle, |gee\rangle,|egg\rangle$ and two antiferromagnetic symmetric (AFMS) states $|geg\rangle, |ege\rangle$~\cite{exp}. However, the competition between $G_{12}$ and $G_{13}$ will break the symmetry in the AFM order. For instance when $G_{13}>G_{12}>0$, the next nearest neighbor interaction wins so that the spin ground state is the superposition of four degenerate AFMA states.

As we discussed in Sec. III, the relative strengths and signs of the couplings $G_{ij}$ are controlled by the modulation frequency $\omega$ and as we can see  from Fig.~\ref{fig:Couplings}, this allows us to access several distinct parameter regimes depending on the relation between $G_{12}$ and $G_{13}$. In the following we denote by region I, II and III  the tree different regimes for $\omega$ which are separated by the  two phonon mode frequencies $\omega_{2}$ and $\omega_{3}$. More precisely, for our model to be valid, $\omega$ can not be too close to resonance and has to satisfy the condition $|\omega-\omega_n|\gg g_{n,i}/2$ for any $i$ and $n$. In the following we set $\epsilon=0.1$ and restrict the detunings to $|\omega-\omega_n|\geq 10\times {\rm max}\{g_{n,i}/2\}$. Then the valid regions for $\omega$ are: region I: $0<\omega<1.84\nu$, region II: $2.63\nu<\omega<3.23\nu$ and region III: $\omega>3.78\nu$. In region I a change of the sign of both coupling constants occurs at $\omega=1.35\nu$  and therefore we distinguish further between region Ia and Ib.
Table~\ref{tab:spin-order} summarizes the ground states and spin order of the three molecules system in the five different parameter regions in the limit $B^x\rightarrow0$.  Since the parameters $G_{ij}$ change continuously in allowed regions, and the spin order changes at the points of $G_{12}=0$ (region I) and $G_{12}=G_{13}$ (region II), conventional `phase transitions' happen even when $B^x=0$. When $\omega$ crosses a phonon mode (except the COM mode), $G_{ij}$ change discontinuously so that the symmetry of the spins jumps as expected between different regions~\cite{Lin}. However, these transitions are not accessible in this setting due to detuning requirements mentioned above.

\begin{figure}
\centering
\includegraphics[width=0.95\linewidth]{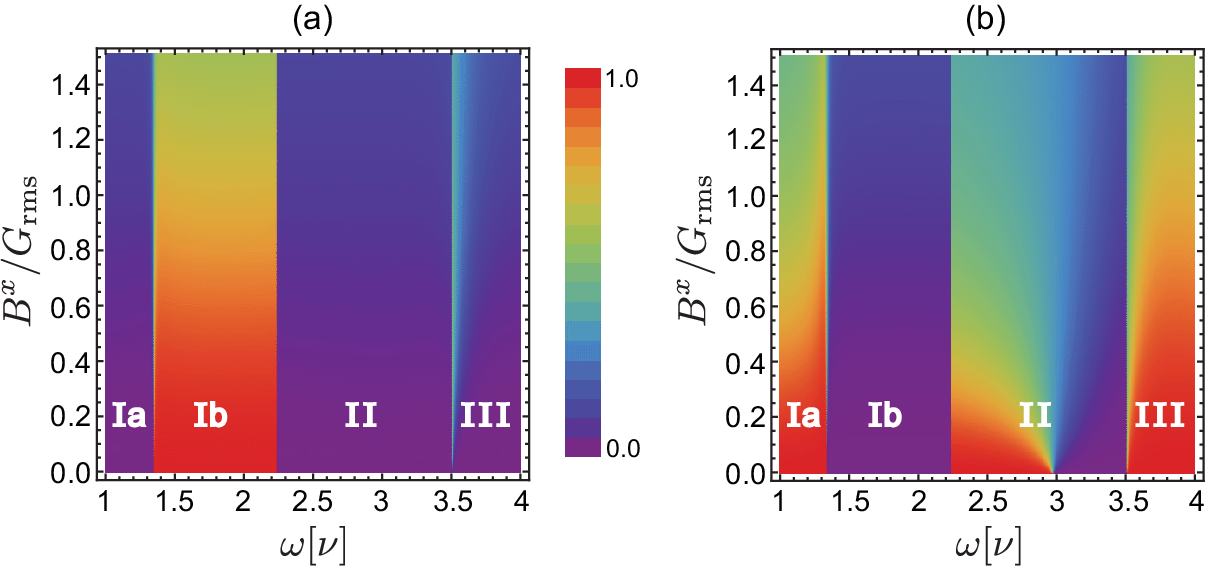}
\caption{(Color online) Ground state of the Ising model~\eqref{eq:HIsing} implemented with three molecules in a harmonic trap. The different regions I, II and III for the modulation frequency $\omega$ as defined in the text are separated by the phonon mode frequencies. (a) Phase diagram showing the overlap of the ground state with the two FM spin configurations, $P_{ggg}+P_{eee}$, as a function of $B^x/G_{\rm rms}$ where $G_{\rm rms}=\sqrt{(G_{12}^2+G_{13}^2+G_{23}^2)/3}$.  (b) Phase diagram showing the overlap of the ground state with the two AFMS spin configuration, $P_{ege}+P_{geg}$. }
\label{fig:phase-diagram}
\end{figure}

Fig.~\ref{fig:phase-diagram} summarize the `phase diagram' of the three molecule system for a finite $B^x$.  To characterize the ground state we follow the approach used in Ref.~\cite{exp}  and plot in Fig. 4(a) the probability of both FM spin combinations $P_{ggg}+P_{eee}$. We see that when $B^x=0$ the spin system is FM ordered in region Ib and there is a phase transition between AFM order and FM order in region I. To distinguish between AFMS and AFMA ground states we plot in Fig. 4(b) the population of two AFMS ground states, $P_{ege}+P_{geg}$. From this plot we can clearly identify a transition between AFMS order and FM order in region I and a  transition between AFMS and AFMA order in region II as indicated in Table~\ref{tab:spin-order}.

\subsection{Adiabatic preparation of entangled ground states}

To see the change in the spin order more directly, for instance the spin order changeing from AFMS to AFMA in region II, we can look at the time evolution of the spin system when the initial state is $|---\rangle$  where $|-\rangle=(|g\rangle-|e\rangle)/\sqrt{2}$, which is the paramagnetic ground state for $B^x\gg G_{\rm rms}=\sqrt{(G_{12}^2+G_{13}^2+G_{23}^2)/3}$. By adiabatically  reducing $B^x/G_{\rm rms}$ the spin system will remain in the ground state of Eq.~\eqref{eq:HIsing}. In Fig. 5 (a) and (b) we numerically evaluate the resulting time evolution for $\omega=2.65\nu$ and $\omega=3.2\nu$ respectively and  plot the overlap of the evolved initial state with an equal superposition of two AFMS states,
\begin{equation}
|\mathrm{AFMS}\rangle=(|geg\rangle-|ege\rangle)/\sqrt2,
\end{equation}
as well as the equal superposition of four AFMA states,
\begin{eqnarray}
|\mathrm{AFMA}\rangle=(|gge\rangle+|egg\rangle-|eeg\rangle-|gee\rangle)/2.
\end{eqnarray}
We see that when  $B^x/G_{rms}\simeq0$, the population of the state $|\mathrm{AFMS}\rangle$ is almost equal to 1 and the spin system shows  AFMS order, while if the system in AFMA order, the final state would be $|\mathrm{AFMA}\rangle$.

\begin{figure}
\centering
\includegraphics[width=0.475\linewidth]{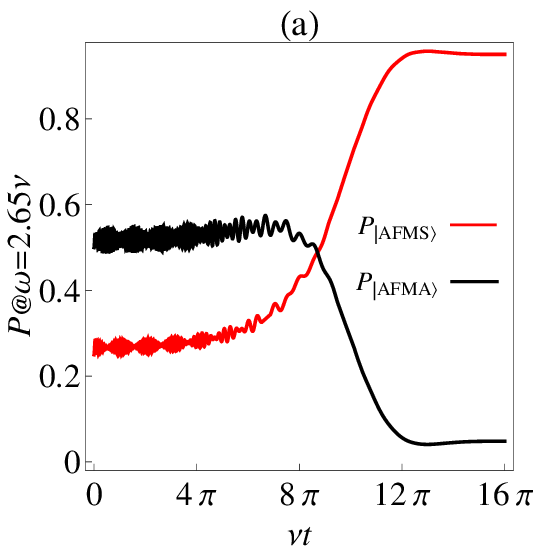}
\includegraphics[width=0.475\linewidth]{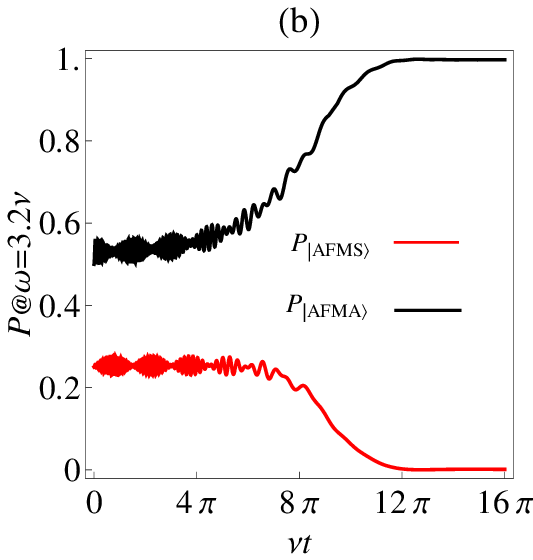}
\includegraphics[width=0.475\linewidth]{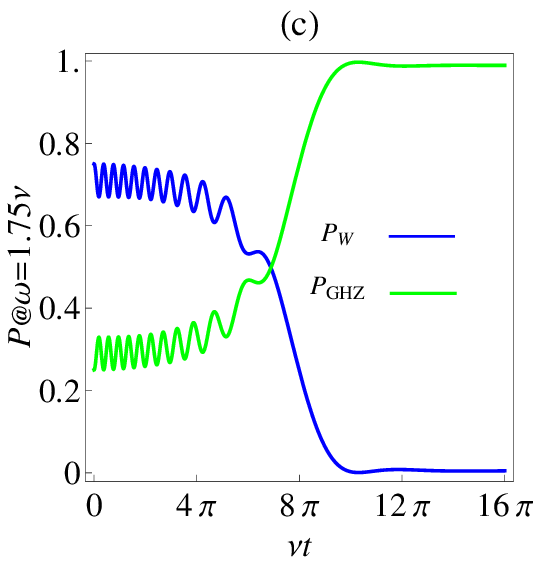}
\includegraphics[width=0.475\linewidth]{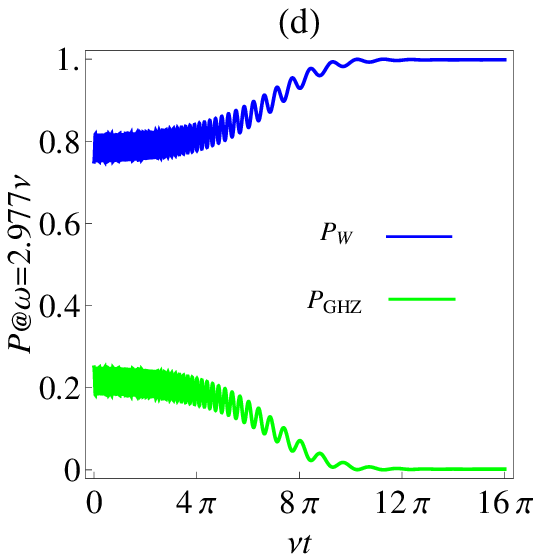}
\caption{(Color online) Adiabatic preparation of the ground state of the three spin Ising model starting from an initial state $|\psi(0)\rangle=| - - - \rangle$ and reducing the transverse field according to $B^x(t)/G_{\rm rms}=10\exp(-(\nu t)^2/(50\pi))$.  The plots (a) and (b) show the overlap of $|\psi(t)\rangle$ with the states $|\mathrm{AFMS}\rangle$ and $|\mathrm{AFMA}\rangle$ for the values of   $\omega=2.65\nu$ and $\omega=3.2\nu$, respectively. The plots (c) and (d) show the overlap of $|\psi(t)\rangle$ with the entangled GHZ and $W$ states as defined in the text.  For these two plots the coupling parameters corresponding to $\omega=1.75\nu$ (c) and $\omega=2.977\nu$ (d) have been assumed.}
 \label{fig:TimeEvolution}
\end{figure}

While so far we have consider the adiabatic time evolution as an alternative way to identify the ground state phases of this system, such an experiment can also be of practical interest when the resulting ground state is an entangled state. From the discussion above we see that when we initialize each spin parallel to a strong transverse field and then adiabatically lower the field compared to  the Ising couplings, the adiabatic evolution of the initial state will result in an equal superposition of all Ising ground states and should therefore be entangled. For instance, in region Ib, the spins show FM order and for a small but non-zero $B^x$ the actual ground state would be the GHZ state $|\mathrm{GHZ}\rangle=(|ggg\rangle-|eee\rangle)/\sqrt2$. For the case of $G_{12}=G_{13}>0$, the resulting ground state would be the equal superposition of six AFM states, which is a superposition of two W state, $W=(|gge\rangle+|egg\rangle+|geg\rangle-|eeg\rangle-|gee\rangle-|ege\rangle)/\sqrt6$ \cite{qs frustrated ising}. The adiabatic preparation of these two special entangled states is summarized in Fig.~\ref{fig:TimeEvolution} where in (c) we have set $\omega=1.75\nu$ to prepare the GHZ state and in (d) $\omega=2.977\nu$ to evolve into a W superposition state.



\section{Scalability}
In our discussion so far we have focused on a system of three molecule in a harmonic trap as the conceptually and experimentally simplest system where frustration phenomena can occur.
In particular, a direct comparison with the analogous trapped ions systems~\cite{exp} shows that all the relevant parameter regimes of the frustrated Ising model are equally well accessible with polar molecules using a simple oscillation electric field.  However, also in analogy to trapped ion systems our techniques relies on a frequency selective addressing of individual phonon modes and is limited to molecular crystals of finite size. Nevertheless, as we discuss in the following, our technique still allows us to go beyond
%
%
proof-of principle experiments and to simulate systems of a few tens of spins, which are no longer trackable using classical simulations.

To proceed let us now for simplicity consider a homogenous 1D system, which could be realized, e.g. using a tight ring trap. This system has periodic boundary conditions, and a constant lattice spacing $a$. The phonon spectrum can then be obtained in a closed form~\cite{rabl-pra} and be written as
\begin{eqnarray}
\hbar\omega_n=\sqrt{\frac{1}{\gamma}}\frac{D}{a^3}\tilde \omega_n,
\end{eqnarray}
where to a good approximation $\tilde \omega_n\simeq 2\sqrt 12 |\sin(\pi n/N)|$ is the dimensionless phonon frequency and $\gamma=Dm/(\hbar^2a)\gg1$ is the ratio between potential and kinetic energy. Note that in contrast to the harmonic trap, here we assume mode labels $n$ from $-N/2$ to $N/2$ and modes with identical $|n|$ are degenerate.  The maximum frequency of the phonon spectrum is the Debye frequency which in units of the dipole-dipole interaction $D/(\hbar a^3)$ is given by $\bar \omega_D\equiv\bar \omega_{n=N/2}\simeq6.94 /\sqrt{\gamma}$ with $\bar\omega_n=\tilde\omega_n/\sqrt\gamma$. The lowest frequency corresponding to the rotation of the whole ring is zero and doesn't play a role in the following discussion. We can write the normalized position fluctuations $\delta\bar x_j=\delta x_j/a$ as
\begin{eqnarray}
\delta\bar x_j=\frac{1}{\sqrt N}\sum_{n}\sqrt{\frac{1}{2\bar \omega_n \gamma}}\left(a_n e^{i2\pi j n/N}+a_{n}^\dagger e^{-i2\pi j n /N}\right),
\end{eqnarray}
from which we obtain a normalized spin-phonon-coupling
\begin{equation}
\bar g_{n,j}= -i   \sqrt{\frac{\epsilon^2}{2N \gamma\bar \omega_n}  } e^{i2\pi j n /N} \tilde g_n,
\end{equation}
where  $\tilde g_n\simeq 6\sin(2\pi n/N)$. Then the bare coupling $\bar{G}_{ij}^0/2 = 1/(2|i - j|^3)$ and by defining $\tilde\omega=\sqrt\gamma\omega/(D/\hbar a^3)$ the phonon-mediated spin-spin coupling can be written as
\begin{eqnarray}
\frac{\bar G_{ij}^1}{2}=\sum_{n>0}\frac{\tilde g_n^2\cos[2\pi(i-j)n/N]}{N(\tilde\omega^2-\tilde\omega_n^2)}.
\end{eqnarray}
 Note that the resulting phonon mediated couplings are independent of $\gamma$, but a value of $\gamma \gg1 $ has to be assumed for a (quasi) crystalline phase and the validity of our model.

\begin{figure}
\includegraphics[scale=0.9]{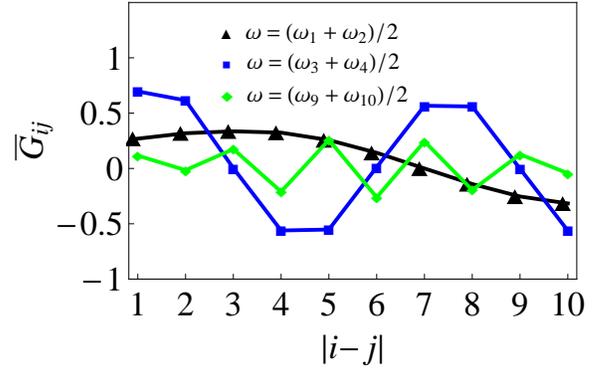}
\caption{(Color online)  The normalized spin-spin coupling strength $\bar G_{ij}$ for $N=21$ polar molecules in a ring trap as a function of the distance when we fix the dipole moment frequency at $(\omega_1+\omega_2)/2$, $(\omega_3+\omega_4)/2$ and $(\omega_9+\omega_{10})/2$, respectively.}
\label{fig:Gij}
\end{figure}

In Fig.~\ref{fig:Gij} we plot the resulting total spin-spin couplings $\bar G_{ij}$ for the case of $N=21$ and different choices of the modulation frequency $\omega$. For a low frequency $\omega=(\omega_1+\omega_2)/2$ we obtain slowly varying, long-range interaction. Since the coupling of each spin to several nearby spins is positive and of similar strength, spin frustration effects as described above will also lead in this configuration to highly entangled grounds states.  This is also true for the second example at $\omega=(\omega_3+\omega_4)/2$ where the coupling is oscillating more rapidly, but still exhibits the properties $G_{i,i+1}\approx G_{i,i+2}>0$ as well as $\sum_{j\neq i} G_{ij}>0$, which prevents a simple anti-ferromagnetic or ferromagnetic ordering. Finally, the last example $\omega=(\omega_{9}+\omega_{10})/2$ shows a case where we address a high frequency phonon mode. Here the coupling exhibits site-to-site oscillations and we can expect an anti-ferromagnetic alignment of spins in the ground state. Once the modulation frequency is tuned above $\omega_D$, the couplings start to decay faster until only the bare coupling $G_{ij} \approx G_{ij}^0/2$ is left for detunings much larger than the mode spacing. Note that while in these three examples the total coupling to other spins $\sum_{j\neq i} G_{ij}$ is positive, also to opposite regime could be achieved by slightly detuning $\omega$ closer to one of the phonon modes. This would allow to tune between ferro- and anti-ferromagnetic long-range couplings.


We now estimate the maximal number of spins that we could control by our technique. The validity of our model requires first of all that the effective spin dynamics is slow compared to the minimal detuning, i.e. $|G_{ij}/\hbar | < {\rm min}\{ |\Delta_n=\omega_n -\omega|\}$. This can in principle always be achieved by choosing a lower value for $\epsilon$, but restricts the achievable spin-spin interactions to about $\sim D/(Na^3)$.  For a nearest neighbor dipole-dipole interaction strength of about 50 kHz and typical coherence times in optical lattice experiments in the range of 100 ms, this would still allow the simulation spin systems of $N\sim100$ spins.

A second condition for our model to be valid is that the maximal relative displacement $\Delta x= {\rm max}\{ |x_i-x_{i+1}|\}$ between the molecules remains small compared to the lattice spacing $a$. To evaluate if this condition can be satisfied, it is convenient to switch to a frame rotating with the modulation frequency $\omega$ and change into an interaction picture with respect to $\sum_n\hbar\omega a_n^{\dagger}a_n$. Then the relevant slowly varying part of Eq.~\eqref{eq:Hrotframe}  becomes
\begin{equation}
H=\sum_n \hbar\Delta_n a_n^{\dagger}a_n
+\sum_{n,i}\frac{\hbar}{2}( g_{n,i} a_n+  g_{n,i}^*a_n^{\dagger})\sigma_i^z+\sum_{i<j}\frac{\hbar G_{ij}^0}{2}\sigma_i^z\sigma_j^z,
\end{equation}
In this form the Hamiltonian can be diagonalize by the unitary displacement operator $U=\exp(\sum_n\sigma_i^z( g_{n,i}^* a_n^{\dagger}- g_{n,i} a_n)/2\Delta_n)$~\cite{polaron-trans}
which displaces the phonon modes by $a_n\rightarrow a_n -\sum_j   g_{n,j}^*\sigma_z^j/(2\Delta_n)$, and correspondingly shifts the position of each molecule by
\begin{equation}
\delta \bar x_i \rightarrow \delta\bar x_i  +   \sum_{n=1}^{N/2} \frac{\epsilon}{\tilde \omega_n}  \frac{\tilde g_{n}}{\tilde\Delta_n}  \frac{1}{N}   \sum_j  \sin\left(\frac{2\pi (j-i) n}{N}\right)   \sigma_z^j .
\end{equation}
Since the system is homogenous we can chose $i=0$ and evaluate the relative displacement $|\delta\bar x_0-\delta\bar x_{1}|$. Assuming that the main contributions come from the lower part of the phonon spectrum we  can write the result approximately as
 \begin{equation}
|\delta\bar x_0-\delta\bar x_{1}| \simeq  \Big|  \sum_{n=1}^{N/2} \frac{\epsilon}{\tilde \omega_n }  \frac{\tilde g_{n}}{\tilde\Delta_n}  \frac{2\pi  n}{N}    S_n  \Big|,
\end{equation}
where $S_n =   \frac{1}{N}   \sum_j  \cos\left(\frac{2\pi j  n}{N}\right)    \sigma_z^j  \leq 1$. The relative displacements depend on the spin configuration $S_n$. For a purely ferro- or anti-ferromagnetic ordering only $S_{n=0}$ or $S_{n=N/2}$ are non-zero, but for both modes $\tilde g_{n}=0$. If we assume a completely random distribution of spin states  we can use the bound $|S_n| \leq 1/\sqrt{N}$ and as above we consider a modulation frequency  $\tilde\omega=(\tilde \omega_{n_0}+\tilde \omega_{n_0+1})/2$ between two phonon modes. Then, apart from a numerical constant $\mathcal{O}(1)$ we obtain the scaling $|\delta\bar x_0-\delta\bar x_{1}| \sim \epsilon \sqrt{N}$. Therefore, compared to the adiabatic condition mentioned above the assumption of small relative displacements does not impose any additional restrictions.

\section{molecular spin qubits}

We finally come back to one of the initial assumptions made in our derivation and discuss in more detail, how to choose an appropriate encoding of the spin states $|g\rangle$ and $|e\rangle$  to obtain the correct form of the induced dipole moments $\mu_{g,e}(t)$ as shown in Fig.~\ref{fig:setup} (b). In the following we illustrate a basic encoding scheme for the case of a molecule with a $^2\Sigma$ ground state~\cite{2sigma-molecules}, where states $|g\rangle$ and $|e\rangle$ are represented in two Zeeman sublevels of two excited rotational states. However, we point out that recently several other schemes for tunable dipole-dipole interactions for spin and/or rotational degrees of freedom~\cite{michael, schachenmayer,Gorshkov} have been suggested which could also be adapted for the present purpose.

\begin{figure}
\includegraphics[scale=0.6]{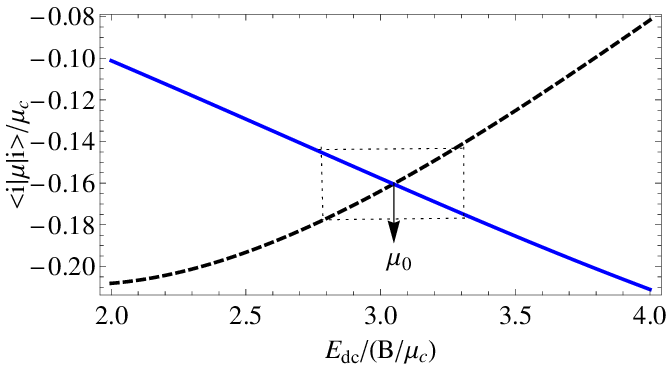}
\caption{(Color online) Induced dipole moments $\langle i|\mu_z|i\rangle$ for $|1,0\rangle_{E_{\mathrm{dc}}}$ (dashed line) and $|2,0\rangle_{E_{\mathrm{dc}}}$ (solid line) as a function of the applied dc field $E_{\mathrm{dc}}$. At the point $E_{\mathrm{dc}}/(B/\mu_c)=3.05$, those two states have the same induced dipole moment $\mu_0$. In the range of rectangle, the dipole moment is approximately linear with external field, and the derivative of the dipole moments of two spin-states with respect to external field is the same but with different signs.}
\label{fig:induce-dipole}
\end{figure}

We consider a $^2\Sigma$ molecule in the vibrational ground state with a single unpaired electron with spin $S=1/2$ and no nuclear spin. In the presence of a external field $\mathbf{E}=E_{dc}\mathbf{e}_z$ the rotational spectrum is described by the Hamiltonian
\begin{eqnarray}\label{eq:Hrot}
H_M=B\mathbf{J}^2-\mathbf{\mu}_c\cdot \mathbf{E}+\gamma_{\mathrm{rs}} \mathbf{J}\cdot\mathbf{S}.
\end{eqnarray}
The first term is the rigid rotor Hamiltonian with $B$ the rotational constant which is typically in the order of several GHz, and $\mathbf{J}$ is the total angular momentum. We denote the energy eigenstates of this rotor Hamiltonian by $|J,M\rangle$ with the spectrum $E_J=BJ(J+1)$, where $M$ is the quantum number of associated with $J_z$. The second term is the Stark interaction which occurs when an electric field is applied to a molecule possessing a permanent electric dipole moment $\mu_c$. The external field mixes different rotational states and splits the $(2J+1)$-fold degeneracy in the rotor spectrum, which amounts to inducing a finite dipole moment in each rotational state. We denote the rotational eigenstates in the presence of the bias field as $|J,M\rangle_{E_{\mathrm{dc}}}$. The third term in Eq.~\eqref{eq:Hrot} is the spin-rotation interaction with $\gamma_{\mathrm{rs}}$ is the spin-rotation coupling constant.  Typically $\gamma_{\mathrm{rs}}\sim 100$ MHz $\ll B$.

For moderate and strong electric fields $E_{\mathrm{dc}}\gg B/\mu_c$, the Stark interaction typically exceeds the spin-rotation coupling and eigenstates of $H_{M}$ are approximately given by product states $|J,M\rangle_{E_{\mathrm{dc}}}\otimes|M_S\rangle$ \cite{rabl-pra} with $M_S=\pm1/2$. Therefore, the dipole moment of each eigenstate is $\langle M_S|\otimes\ _{E_{\mathrm{dc}}}\langle J, M|\mu_z|J, M\rangle_{E_{\mathrm{dc}}}\otimes|M_S\rangle\simeq\ _{E_{dc}}\langle J, M|\mu_z|J, M\rangle_{E_{\mathrm{dc}}}$. We encode our effective spin system in the two states $|g\rangle=|1,0\rangle_{E_{\mathrm{dc}}}\otimes|-1/2\rangle$ and $|e\rangle=|2,0\rangle_{E_{\mathrm{dc}}}\otimes|1/2\rangle$, so that $\mu_e=\langle e|\mu_z|e\rangle=\ _{E_{\mathrm{dc}}}\langle 2,0|\mu_z|2,0\rangle_{E_{\mathrm{dc}}}$ and $\mu_g=\langle g|\mu_z|g\rangle=\ _{E_{\mathrm{dc}}}\langle1,0|\mu_z|1, 0\rangle_{E_{\mathrm{dc}}}$. For the spin-forbidden transition, transition matrix element between $|g\rangle$ and $|e\rangle$ are small \cite{rabl-pra} and compared with the induced dipole moments, the transition dipole moments $\langle e|\mu_z|g\rangle$ are suppressed by the ratio $(\gamma_{\mathrm{rs}}/B)^2$.  However, the dipole-dipole interaction can still induce a spin preserving flip-flop process where the state of first molecule changes as  $|1,0\rangle \otimes |-1/2\rangle \rightarrow  |2,0\rangle \otimes |-1/2\rangle$ and the state of the second molecule as $|2,0\rangle \otimes |1/2\rangle \rightarrow  |1,0\rangle \otimes |1/2\rangle$ and brings the molecules out of the initial two level subspace. These processes can be suppressed by energetically shifting e.g. the state  $|2,0\rangle \otimes |-/2\rangle \rightarrow  |1,0\rangle \otimes |1/2\rangle$ by coupling it off-resonantly  with a circularly polarized microwave field to the $J=3$ manifold.

In Fig.~\ref{fig:induce-dipole} we plot the induced dipole moments for two rotational states $|1,0\rangle_{E_{\mathrm{dc}}}$ and $|2,0\rangle_{E_{\mathrm{dc}}}$ as a function of the external electric field $E_{\mathrm{dc}}$. At the ``sweet spot" $E_{\mathrm{ac}}=E_0=3.05(B/\mu_c)$, the induced dipole moments for those two states are both equal to $\mu_0=-0.16\mu_c$. In the range indicated by the rectangle in Fig.~\ref{fig:induce-dipole}, we can make two approximations about spin-states $|g\rangle$ and $|e\rangle$. i) Their dipole moments are approximately linear with the electric field strength; ii) The derivative of the dipole moments of two spin-states with respect to external field is approximately the same, but with different signs, i.e., $\partial_{E_{\mathrm{dc}}}\mu_{g}|_{E_0}\approx -\partial_{E_{\mathrm{dc}}}\mu_{e}|_{E_0}=\partial_{E_{dc}}\mu|_{E_0}$. These conditions imply that when we fix the dc bias field at the sweet spot and add another time dependent weak field $E_{\mathrm{ac}}(t)=E_{\mathrm{ac}}\cos(\omega t)$, the dipole moments of two spin states become $\mu(t)=\mu_0+\delta\mu(t)\sigma^z$ with $\delta\mu(t)=\partial_{E_{\mathrm{dc}}}\mu|_{E_0}E_{\mathrm{ac}}\cos(\omega t)$. Note that within the linear regime of the dipole moments variations of $\epsilon =|\delta \mu|/\mu_0\approx 0.1$ as required for our model can be achieved.

\section{Conclusion and outlook}
In this work we have studied the interplay between direct and phonon-mediated spin-spin interactions in dipolar crystals of polar molecules. In particular we have shown that for an appropriate choice of rotational states an electric bias field with time varying amplitude can be used to induce a state dependent modulation of the dipole moments of the molecules, which can resonantly enhance the coupling of the internal degrees of freedom to a specific collective phonon mode in the crystal.
This minimal level of control already enables the implementation of  Ising models with non-trivial spin-spin interactions which for small and moderate sized systems can be used to generate various multi-particle entangled states or to simulate Ising models with `infinite-range' interactions.  To extend this method to larger spin systems and to achieve a more flexible control over the resulting spin-spin interaction patterns different generalizations of our technique can be envisioned. This includes arrays of microtraps where within each trap a small number of molecules interact via phonon-mediated couplings while couplings between traps are achieved via direct dipole-dipole interactions. In this case the frequency splitting between phonon-modes does not decrease with the total systems size.  Further, in analogy to Ref.~\cite{michael}, local defects or so-called `marker' molecules can be introduced to modify the local phonon structure and by that tune between long- and short-range interactions.

%

\section{Acknowledgments}
The authors thank P. Zoller for originally proposing the mechanism behind the effective spin-spin interactions discussed in the present paper. This work was supported by AFOSR, MURI, the Austrian
Science Foundation (FWF) through SFB F40 FOQUS and by the EU Networks NAMEQUAM and AQUTE.
Y. L. Zhou acknowledges support by Hunan Provincial Innovation Foundation For Postgraduates and NSFC grant No. 11074307.

\appendix
\section{Effective spin dynamics and state purity}\label{app}
In the absence of any transverse magnetic fields the full time dependent Hamiltonian given in Eq.~\eqref{eq:Hrotframe} can be integrated exactly and
 the resulting time evolution operator can be written as
\begin{equation}
U(t)= e^{- i \sum_{n,i}  ( \alpha_{n,i}(t) a_n +  \alpha^*_{n,i}(t) a_n^\dag) \sigma_z^i} e^{-i \sum_{i<j}Ê\Phi_{ij}(t) \sigma_z^i\sigma_z^j}.
\end{equation}
Here the displacement amplitudes are
\begin{equation}
\alpha_{n,i}= i \frac{g_{n,i}}{2}  \left[ \frac{1-e^{-i(\omega_n-\omega)t}}{\omega-\omega_n} - \frac{1-e^{-i(\omega_n+\omega)t}}{\omega+\omega_n} \right],
\end{equation}
and the phases $\Phi_{ij}(t)=\Phi_{ij}^0(t)+ \Phi_{ij}^1(t)$ are given by the bare  coupling
\begin{equation}
\Phi^0_{ij}(t)=  \frac{G_{ij}^0}{2} \left( t +\frac{\sin(2\omega t)}{2\omega}\right),
\end{equation}
and the phonon-mediated part
\begin{equation}
\begin{split}
\Phi^1_{ij}(t)= &\sum_{n}  \frac{2g_{n,i} g_{n,j}\omega_n}{(\omega^2-\omega_n^2)}  \Big[ \frac{t}{2} +\frac{\sin(2\omega t)}{4\omega}\\
&+ \frac{ \omega_n \cos(\omega t)\sin(\omega_n t) + \omega \sin(\omega t)\cos(\omega_n t)}{(\omega^2-\omega_n^2)}\Big].
\end{split}
\end{equation}
From these expressions we can extract the slowly varying spin-spin couplings $G_{ij}^{0,1} := \lim_{t\rightarrow \infty}Ê\Phi_{ij}^{0,1}/t$.

To evaluate the purity of the spin subsystem during the evolution we write an arbitrary initial spin superposition state as $|\Psi_0\rangle= \sum_{\vec s}  C_{\vec s} |\vec s\rangle
$ where $\sigma_z^i|\vec s\rangle =s_i|\vec s\rangle$. With $\rho_{ph}$ being the equilibrium density operator of the phonon modes the total density operator at time $t=0$ is $\rho(0)=|\Psi_0\rangle\langle \Psi_0|\otimes \rho_{ph}$ and the reduced spin density operator at time $t$ is given by
\begin{equation}
\rho_s(t) = \sum_{\vec s,\vec r}  C_{\vec s} C^*_{\vec r}  e^{-i \Phi(t,\vec s, \vec r)}  e^{- \frac{1}{2}
 \sum_{n} F_n(t,\vec s,\vec r)}  |\vec s\rangle\langle \vec r|.
\end{equation}
Here $\Phi(t,\vec s, \vec r)=\sum_{i<j}Ê\Phi_{ij}(\vec s_i \vec s_j -\vec r_i \vec r_j)$ and
\begin{equation}
F_n(t,\vec s,\vec r) =  \left[\sum_i \alpha_{n,i}(t) (\vec s_i-\vec r_i)\right]^2 \coth\left(\frac{\hbar \omega_n}{2k_BT}\right),
\end{equation}
where $T$ is the temperature of the phonon modes. The purity $P_s(t)={\rm Tr}\{\rho^2_s(t)\}$ of the spin system can then be written as
\begin{equation}
P_s(t) = \sum_{\vec s,\vec r}  |C_{\vec s}|^2 |C^*_{\vec r}|^2    e^{- \sum_{n} F_n(t)(\vec s,\vec r)}.
\end{equation}


\begin{thebibliography}{99}

\bibitem{quantum-phase-book} S. Sachdev, {\it Quantum Phase Transitions} (Cambridge University Press, Cambridge, 1999).

\bibitem{frustrated-book} J. J. Binney, N. J. Dowrick, A. J. Fisher and M. E. J. Newman, {\it The Theory of Critical Phenomena: An Introduction to the Renormalization Group}  (Oxford University Press, Oxford, 1992).

\bibitem{cluster1} R. Raussendorf, and H. J. Briegel, Phys. Rev. Lett. {\bf 86}, 5188 (2001).
\bibitem{cluster2} H. J. Briegel, and R. Raussendorf, Phys. Rev. Lett. {\bf 86}, 910 (2001).
\bibitem{graph1} M. Hein, J. Eisert, and H. J. Briegel, Phys. Rev. A {\bf 69}, 062311 (2004).

\bibitem{ent-frus} L. Amico, R. Fazio, A. Osterloh and V. Vedral, Rev. Mod. Phys. {\bf 80} (2), 517 (2008). 

\bibitem{qsor} I. Buluta and F. Nori, Science {\bf 326}, 108 (2009). 



\bibitem{HaeffnerPhysRep2008}Ê H. Haeffner, C.F. Roos, and R. Blatt, Phys. Rep. {\bf 469}, 155 (2008).
\bibitem{WinelandLaserPhysics2011} D. J. Wineland and D. Leibfried, Laser Phys. Lett. {\bf 8}, 175 (2011). 


\bibitem{MM gate} K. M{\o}lmer, and A. S{\o}rensen. Phys. Rev. Lett. {\bf 82}, 1835 (1999).
\bibitem{ions-spin} D. Porras, and J. I. Cirac, Phys. Rev. Lett. {\bf 92}, 207901 (2004).
\bibitem{XXZmodel} Philipp Hauke, Fernando M. Cucchietti, Alexander M¨¹ller-Hermes, Mari-Carmen Ba\~{n}uls, J. Ignacio Cirac, Maciej Lewenstein, New J. Phys. {\bf 12}, 113037 (2010).    
\bibitem{IsingModels} J. J. Garc\'ia-Ripoll, P. Zoller and J. I. Cirac, Phys. Rev. A \textbf{71}, 062309 (2005).
\bibitem{deng} X. L. Deng, D. Porras, and J. I. Cirac, Rhys. Rev. A {\bf 72}, 063407 (2005).
\bibitem{Zhu} S.-L. Zhu, C. Monroe, and L. M. Duan, Phys. Rev. Lett. {\bf 97}, 050505 (2006).
\bibitem{Lin} G.-D. Lin, C. Monroe, and L.-M. Duan, Phys. Rev. Lett. {\bf 106}, 230402 (2011).

\bibitem{Friedenauer} A. Friedenauer, H. Schmitz, J. T. Gl\"ukert, D. Porras, T. Sch\"atz, Nature Physics {\bf 4}, 757 (2008).
\bibitem{Kim-prl} K. Kim, M.-S. Chang, R. Islam, S. Korenblit, L.-M. Duan, and C. Monroe, Phys. Rev. Lett. {\bf 103}, 120502 (2009).
\bibitem{exp} E. E. Edwards, S. Korenblit, K. Kim, R. Islam, M. S. Chang, J. K. Freericks, G. D. Lin, L. M. Duan, and C. Monroe, Phys. Rev. B {\bf 82}, 060412 (2010).
\bibitem{qs frustrated ising} K. Kim, M.-S. Chang, S. Korenblit, R. Islam, E. E. Edwards, J. K. Freericks, G.-D. Lin, L.-M. Duan, and C. Monroe, Nature {\bf 465}, 590 (2010).




\bibitem{coolingEXP1}
J. D. Weinstein, R. deCarvalho, T. Guillet, B. Friedrich, and J. M. Doyle, Nature \textbf{359}, 148 (1998).
\bibitem{coolingEXP2}
H. L. Bethlem, G. Berden and G. Meijer, Phys. Rev. Lett. \textbf{83} , 1558 (1999).
\bibitem{coolingEXP3}
A. J. Kerman, J. M. Sage, S. Sainis, T. Bergeman and D. DeMille, Phys. Rev. Lett. \textbf{92}, 033004 (2004).
\bibitem{coolingEXP4} K. K. Ni, S. Ospelkaus, M. H. G. de Miranda, A. Pe'er, B. Neyenhuis, J. J. Zirbel, S. Kotochigova, P. S. Julienne, D. S. Jin and J. Ye, Science \textbf{322}, 231 (2008).
\bibitem{coolingEXP5}
F. Lang, K. Winkler, C. Strauss, R. Grimm and J. HeckerDenschlag, Phys. Rev. Lett. \textbf{101}, 133005 (2008).
\bibitem{coolingEXP6}
J. Deiglmayr, A. Grochola, M. Repp, K. M\"ortlbauer, C. Gl\"uck, J. Lange, O. Dulieu, R. Wester, and M. Weidem\"uller, Phys. Rev. Lett. \textbf{101}, 133004 (2008).
\bibitem{coolingEXP7}
K. Aikawa, D. Akamatsu, J. Kobayashi, M. Ueda, T. Kishimoto and S. Inouye, New J. Phys. \textbf{11}, 055035 (2009).
\bibitem{coolingEXP8}
E. S. Shuman, J. F. Barry and D. DeMille, Nature \textbf{467}, 820 (2010).




\bibitem{revPM1}
J. Doyle, B. Friedrich, R. V. Krems and F. Masnou-Seeuws, Special Issue on Ultracold Polar Molecules: Formation and Collisions, Eur. Phys. J. D. {\bf 31}, 149 (2004).
\bibitem{revPM2}
L. D. Carr, D. DeMille, R. V. Krems and J. Ye, New J. Phys. \textbf{11}, 055049 (2009).

\bibitem{MDC1}
H. P. B\"uchler, E. Demler, M. Lukin, A. Micheli, N.  ProkofÕev, G. Pupillo, and P. Zoller, Phys. Rev. Lett. \textbf{98}, 060404 (2007).
\bibitem{MDC2}
G. E. Astrakharchik, J. Boronat, I. L. Kurbakov and Y. E. Lozovik, Phys. Rev. Lett. \textbf{98}, 060405 (2007).
\bibitem{MDC3}
R. Citro, E. Orignac, S. DePalo and M. L. Chiofalo, Phys. Rev. A \textbf{75}, 051602(R) (2007).
\bibitem{rabl-pra} P. Rabl, and P. Zoller, Phys. Rev. A {\bf 76}, 042308 (2007).

\bibitem{Micheli-Nat} A. Micheli, G. K. Brennen, and P. Zoller, Nature Physics {\bf 2}, 341 (2006).

\bibitem{Micheli2-Nat} H. P. B\"uchler, A. Micheli, and P. Zoller, Nature Physics {\bf 3}, 726 (2007). 


\bibitem{schachenmayer}ÊJ. Schachenmayer, I. Lesanovsky, A. Micheli, A. J. Daley, New J. Phys. {\bf 12} 103044 (2010). 

\bibitem{HerreraPRA}  F. Herrera, M. Litinskaya, and R. V. Krems,  Phys. Rev. A {\bf 82}, 033428 (2010). 

\bibitem{TrefzgerNJP2010}  C. Trefzger, M. Alloing, C. Menotti, F. Dubin and M. Lewenstein, New J. Phys. {\bf 12}, 093008 (2010). 

\bibitem{WallPRA2010} M. L. Wall and L. D. Carr, Phys. Rev. A {\bf 82}, 013611 (2010). 

 \bibitem{KestnerPRB2010} J. P. Kestner, Bin Wang, Jay D. Sau, and S. Das Sarma,  Phys. Rev. B {\bf 83}, 174409 (2011). 

\bibitem{Gorshkov}  A. V. Gorshkov, S. R. Manmana, G. Chen, J. Ye, E. Demler, M. D. Lukin, A. M. Rey, Phys. Rev. Lett. {\bf 107}, 115301 (2011); A. V. Gorshkov, S. R. Manmana, G. Chen, E. Demler, M. D. Lukin, A. M. Rey, arXiv:1106.1655 (2011).







\bibitem{michael} M. Ortner, Y. L. Zhou, P. Rabl, P. Zoller, arXiv: quant-ph/1106.0128v2 (to appear in Quantum Information Processing).

 \bibitem{DeMille} D. DeMille, Phys. Rev. Lett. {\bf 88}, 067901 (2002).
\bibitem{QIPswitch} S. F. Yelin, K. Kirby and R. C\^ot\'e, Phys. Rev. A \textbf{74}, 050301(R) (2006).




%








\bibitem{michael-njp} M. Ortner, A. Micheli, G. Pupillo, and P. Zoller, New J. Phys. {\bf 11}, 055045 (2009).

\bibitem{herrera-arXiv} F. Herrera, R. V. Krems, arXiv:1010.1782 (2010).




 \bibitem{Micheli-pra} A. Micheli, G. Pupillo, H. P. B\"uchler, and P. Zoller, Phys Rev A 76, 043604 (2007).



\bibitem{atom-spin} M. J. Hartmann, F. G. S. L. Brandao and M. B. Plenio, Phys. Rev. Lett. {\bf 99}, 160501 (2007).
\bibitem{James} G. J. Milburn, S. Schneider, and D. F. V. James. Fortschr. Phys. {\bf 48}, 801 (2000).

\bibitem{AndreNatPhys2006}ÊA. Andre, D. DeMille, J. M. Doyle, M. D. Lukin, S. E. Maxwell, P. Rabl, R. J. Schoelkopf and P. Zoller,
Nature Phys. {\bf 2}, 636 (2006).

\bibitem{loewen} M. Loewen, and C. Wunderlich, Verhandl. DPG VI 39/113 (2004).

\bibitem{polaron-trans} C. Wunderlich, H. Figger, D. Meschede and C. Zimmermann, {\it Laser Physics at the Limit} (Springer, Heidelberg, 2002).

\bibitem{2sigma-molecules} S. Kotochigova, J. Klos, A. Petrov, M. Linnik, P. S. Julienne, arXiv:1108.3530 (2011).



\end{thebibliography}
\end{document}